\documentclass[aps,pre,twocolumn,letter,superscriptaddress]{revtex4-1}
\usepackage{graphicx}
\usepackage{float}
\usepackage{amsmath}
\usepackage{color}
\usepackage{xr}
\begin{document}
\title{Echoes of the glass transition in athermal soft spheres}
\date{\today}
\author{Peter K. Morse}
\affiliation{Department of Physics, Syracuse University, Syracuse, NY 13244, USA.}
\affiliation{Department of Physics and Materials Science Institute, University of Oregon, Eugene, Oregon 97403, USA.}
\author{Eric I. Corwin}
\affiliation{Department of Physics and Materials Science Institute, University of Oregon, Eugene, Oregon 97403, USA.}

\begin{abstract}
Recent theoretical advances have led to the creation of a unified phase diagram for the thermal glass and athermal jamming transitions.  This diagram makes clear that, while related, the mode-coupling---or dynamic---glass transition is distinct from the jamming transition, occurring at a finite temperature and significantly lower density than the jamming transition.  Nonetheless, we demonstrate a pre-jamming transition in athermal frictionless spheres which occurs at the same density as the mode-coupling transition and is marked by percolating clusters of locally rigid particles. At this density in both the thermal and athermal systems, individual motions of an extensive number of particles become constrained, such that only collective motion is possible. This transition, which is well below jamming, exactly matches the definition of collective behavior at the dynamical transition of glasses. Thus, we reveal that the genesis of rigidity in both thermal and athermal systems is governed by the same underlying topological transition in their shared configuration space.

\end{abstract}

\maketitle

Window glass and sand piles are both amorphous solids; push them and they'll push back. The rigidity in glasses is a thermal phenomenon mediated by the random fluctuations in particle positions, whereas the rigidity in jammed configurations is the result of enduring contacts. A jammed hard-sphere system, suddenly given thermal excitations, will remain rigid \cite{zhang_thermal_2009}, whereas a hard-sphere glass right at the transition, robbed of all thermal excitations, will lose its rigidity \cite{lubchenko_theory_2015}. In the recently proposed glass transition phase diagram \cite{liu_nonlinear_1998, ikeda_unified_2012, charbonneau_jamming_2015}, the jamming transition occurs on the zero temperature - or infinite pressure - line, whereas the glass transition happens at a finite temperature and pressure. However, even as these two transitions are controlled by the same underlying physics and share the same space of allowed configurations \cite{ohern_jamming_2003, ma_potential_2014}, the glass transition happens at a significantly lower density than the jamming transition \cite{berthier_glass_2009, charbonneau_glass_2011, charbonneau_dimensional_2012, charbonneau_universal_2016}. To understand the common origin of rigidity, we link the thermally stabilized rigidity of glassy systems to a fundamental change in the underlying energy landscape of athermal systems.

The athermal jamming transition is defined by the appearance of global rigidity in which all degrees of freedom are constrained.  This global rigidity may arise as a result of increased packing fraction \cite{liu_nonlinear_1998, pusey_phase_1986, coulais_how_2014}, increased friction \cite{onoda_random_1990, song_phase_2008}, or applied shear \cite{bi_jamming_2011,vinutha_disentangling_2016}.  Below the transition lies the mechanical vacuum in which unconstrained internal motions are possible. However, structural order emerges in this mechanical vacuum as the jamming or glass transitions are approached \cite{xia_structural_2015, morse_geometric_2014, malins_identification_2013}. While previous work has connected a rigidity transition in spin networks on Bethe lattices with the mode coupling transition in spin glasses \cite{sellitto_facilitated_2005}, no such link has been made for athermal systems and structural glasses.

\begin{figure}[h]
\includegraphics[width=1\linewidth]{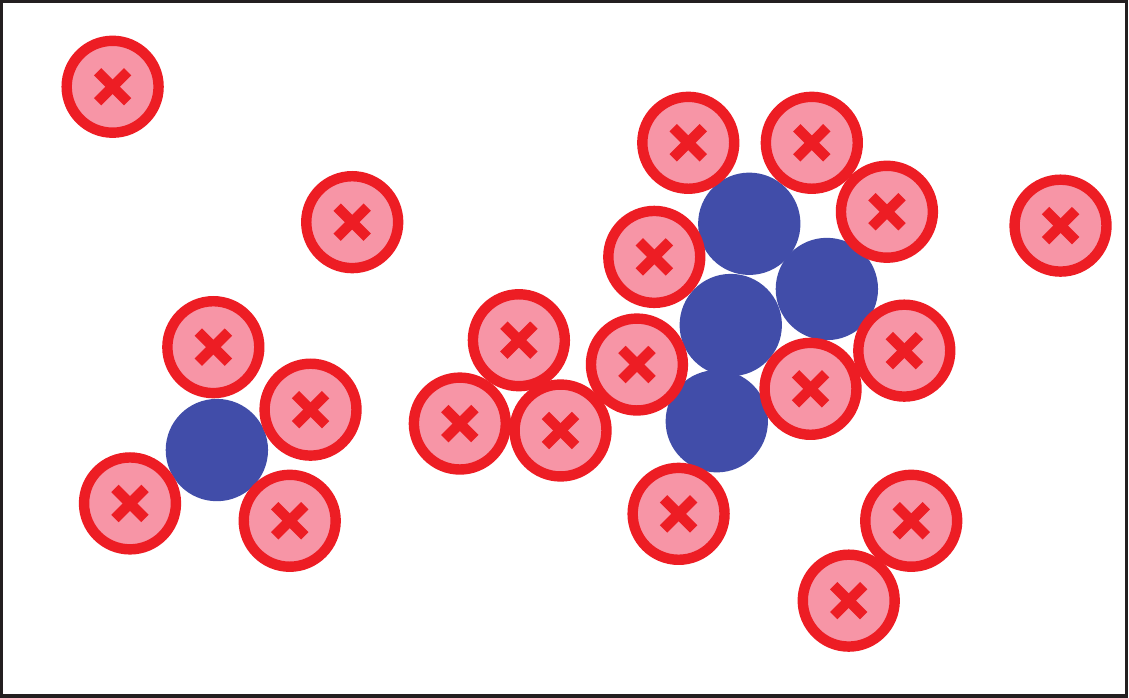}
\caption{Definition of local rigidity: the blue particles are considered locally rigid if all of the red particles marked with an x are constrained. This represents the largest locally rigid set of the system.}
\label{fig:phiLRDef}
\end{figure}

In previous work we have demonstrated the existence of a pre-jamming geometric transition \cite{morse_geometric_2014, morse_hidden_2016} marked by the appearance of particles that are completely hemmed in by a set of kissing nearest neighbors, such that their individual motion is perfectly constrained.  If all other particles are held fixed this particle will be rigid, and such a particle can be called locally rigid. We can exploit this observation by extending it to a collection of particles, which can be called locally rigid if there are no unconstrained internal degrees of freedom within the collection. If all particles outside of the collection are held fixed then all particles in the collection will be rigid, as illustrated in figure \ref{fig:phiLRDef}.  The concept of local rigidity can be expressed in a rigorous fashion by considering the eigenvalues of the Hessian matrix constructed for the system, as described below.

We simulate packings of $N$=256-32768  frictionless athermal monodisperse particles in spatial dimension $d$=3-6 using periodic boundary conditions as described in reference \cite{morse_geometric_2014}.  Our system is composed of monodisperse soft spheres with a harmonic contact potential, making the total energy of the system
\begin{equation}
E = \frac{1}{2}\sum_{i<j}\Theta(\sigma - r_{ij})(\sigma - r_{ij})^2
\end{equation}
where $r_{ij}$ is the distance between particles $i$ and $j$, $\Theta$ is the Heaviside function, and $\sigma$ is the particle diameter. 

We create energy minimized systems at a desired packing fraction $\phi$ using the infinite quench protocol \cite{ohern_random_2002, morse_geometric_2014}.  Particles are placed randomly and uniformly throughout the periodic boundary condition simulation volume.  The particles are given radii such that the system is at the desired packing fraction, and then the energy is minimized.  Because we are below jamming, the zero energy region of configuration space will be extended, analogous to a lake at the bottom of a crater, rather than a single well defined point of local minimum.  To avoid sailing out into this lake of zero energy we modify our conjugate gradient minimization algorithm to stop precisely at the lake shore in this analogy: the edge between zero and non-zero energy.  This is achieved by enforcing that a particle comes to a complete stop as soon as the net force on it drops to zero.  This simple change prevents systems from overshooting into the zero energy sea and ensures that particles which were overlapping before minimization are kissing at the end of the minimization, instead of having a finite space between them. 

As previously stated, we are only considering systems below jamming, so the true energy minimum is always exactly zero. However, our calculations are made with finite precision, and so our zero is taken as some finite but small energy cutoff $E_\textrm{cut}$. Unless otherwise stated, we use $E_\textrm{cut} = N \times 10^{-60}$, roughly the maximum precision achievable with quadruple floating point numbers.  

At a packing fraction of zero no particle motions will be constrained.  Conversely, at the jamming transition we need only restrict one particle to fully constrain the motion of the entire system (excluding rattlers).  Thus all but one non-rattling particles are locally rigid at jamming. Between these extremes there must be a crossover density wherein there is a percolating network of locally rigid particles.

In order to find this crossover density we must make the concept of local rigidity concrete.  We do so by finding the largest locally rigid set. To do this, we first compute the Hessian matrix for a given packing
\begin{equation}
H_{ij}^{\alpha \beta} = \frac{\partial^2E}{\partial r_i^\alpha \partial r_j^\beta}
\end{equation}
where $\alpha$ and $\beta$ represent vector components and $i$ and $j$ particle labels. Eigenvectors and eigenvalues of the Hessian are the vibrational modes and frequencies of the system respectively. Soft modes are ones in which the eigenvalue is equal to zero, corresponding to free motions of particles.  Within this framework, a locally rigid set of particles is one for whom the corresponding submatrix of the Hessian has no zero-energy modes.  The maximum locally rigid set is the largest such set of locally rigid particles possible for a given system, with size $N_\textrm{LR}$.

In practice, finding this largest set is computationally intractable.  However, by counting the number of kissing contacts $N_\textrm{c}$ in the system we can compute the number of zero modes that must be present in the Hessian as $N_0 = (N - 1)d + 1 - N_\textrm{c}$  \cite{maxwell_calculation_1864, goodrich_finite-size_2012,dagois-bohy_soft-sphere_2012}.  Pinning a particle in place is equivalent to removing from the Hessian the rows and columns corresponding to that particle.  If we were to pin in place a single particle participating in a zero mode we would remove between $1$ and $d$ zero modes from the new restricted Hessian.  Thus, a system with $N_0$ zero modes will require that between $N_0/d$ and $N_0$ particles be pinned to find the maximum locally rigid set.  This provides a lower and upper bound: 
\begin{equation}
N - N_0 < N_\textrm{LR} < N- \frac{N_0}{d}.
\end{equation}

The following scheme finds locally rigid sets that nearly saturate the upper bound, with $N_\textrm{LR} \simeq 0.95\times \left( N - N_0/d \right)$.  Any particle without at least $d$ + 1 non co-hemispheric kissing contacts can not be part of a locally rigid set and is immediately pinned.  We then choose a random ordering of the candidate particles and go through this list one-by-one and find the associated submatrix of the Hessian. If adding a new particle introduces a zero mode, we remove that particle from the list and pin it down.  At the end of this process we have identified a list of particles in a nearly largest locally rigid set.  We sample this process many times and note that different random orderings only changes the number of locally rigid particles $N_\textrm{LR}$ by a few particles, even in large systems.

\begin{figure}[htpb]
\includegraphics[width=1\linewidth]{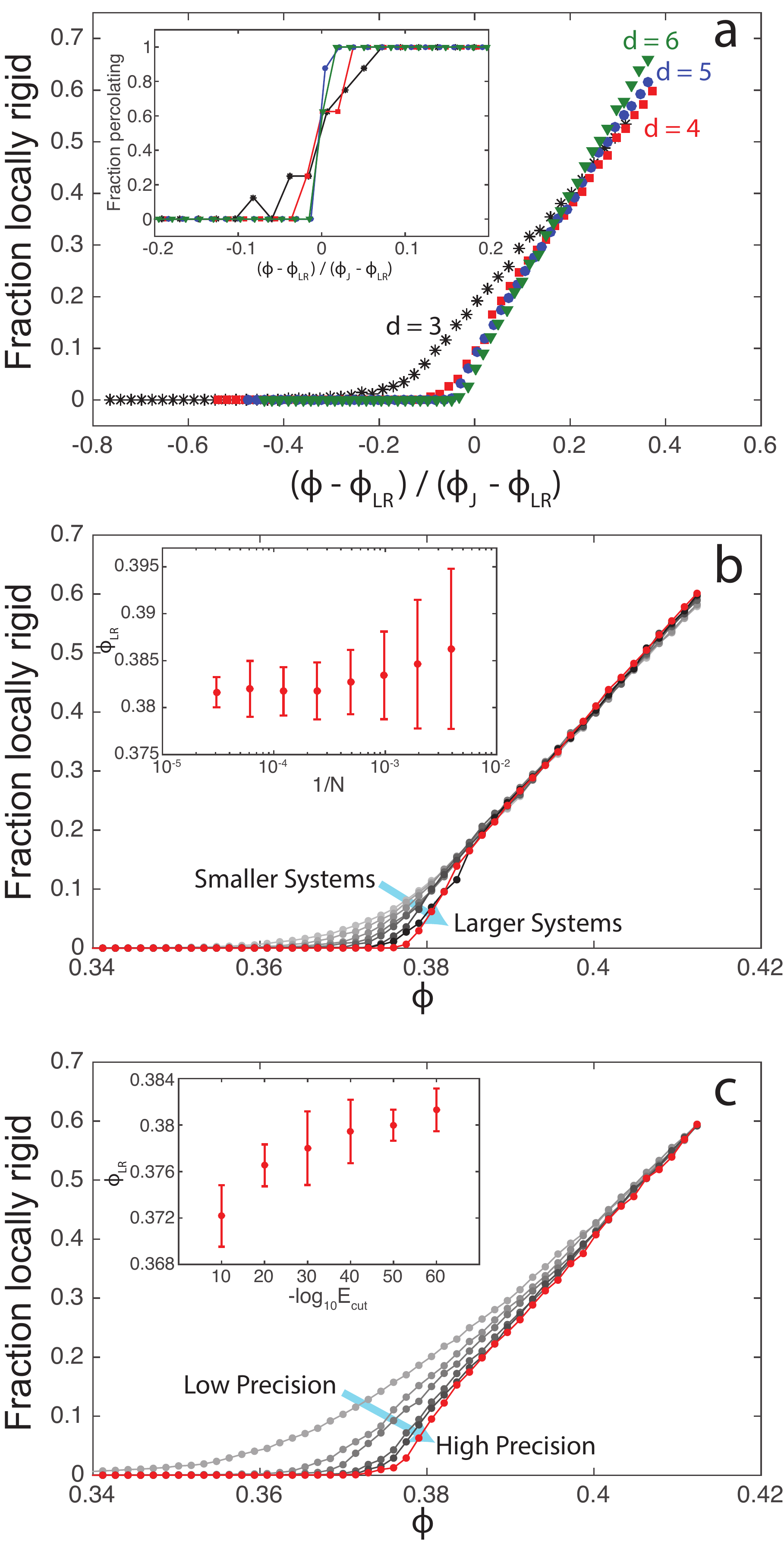}
\caption{The fraction of locally rigid particles as a function of scaled packing fraction. a) For dimensions 3 (black), 4 (red), 5 (blue), and 6 (green) each averaged over 8 systems of $N=16384$ particles. The inset shows the fraction of systems with a percolating locally rigid set. b) Finite size effects in $d=4$ seen for system sizes ranging from N=256 (light gray) to 16384 (black) in powers of 2 and the largest system 32768 in red. We average over 512 systems at N=256, 256 at N=512, 128 at N=1024, 64 at N=2048, 32 at N=4096, 16 at N=8192, 8 at N=16384, and 16 at N=32768. The inset shows the change of $\phi_\textrm{LR}$ as a function of system size. c) Differences in precision in $d=4$ as measured by $-\log_{10}\textrm{E}_\textrm{cut} =$ 10, 20, 30, 40, 50, (light gray to black) and the highest precision of 60 in red. This data is averaged over 16 systems of N=8192 particles. The inset shows how $\phi_\textrm{LR}$ changes with precision.}
\label{fig:phiLRAll}
\end{figure}

\begin{figure}
\begin{tabular}{c|c|c|c|c}
\hspace{0.5cm} d \hspace{0.5cm} & \hspace{0.5cm}3 \hspace{0.5cm} & \hspace{0.5cm} 4\hspace{0.5cm} & \hspace{0.5cm}5\hspace{0.5cm} & \hspace{0.5cm} 6 \hspace{0.5cm} \\
\hline
$\phi_\textrm{LR}$ & 0.55(8) & 0.38(2) & 0.251(0) & 0.159(1)\\
\hline
$\phi_\textrm{d}$ \cite{charbonneau_glass_2011,charbonneau_dimensional_2012} & 0.571 & 0.4065 & 0.2700 & 0.1732\\
\hline
$\phi_\textrm{J}$ \cite{morse_geometric_2014} & 0.6437 & 0.4562 & 0.3079 & 0.1999\\
\end{tabular}
\includegraphics[width=1\linewidth]{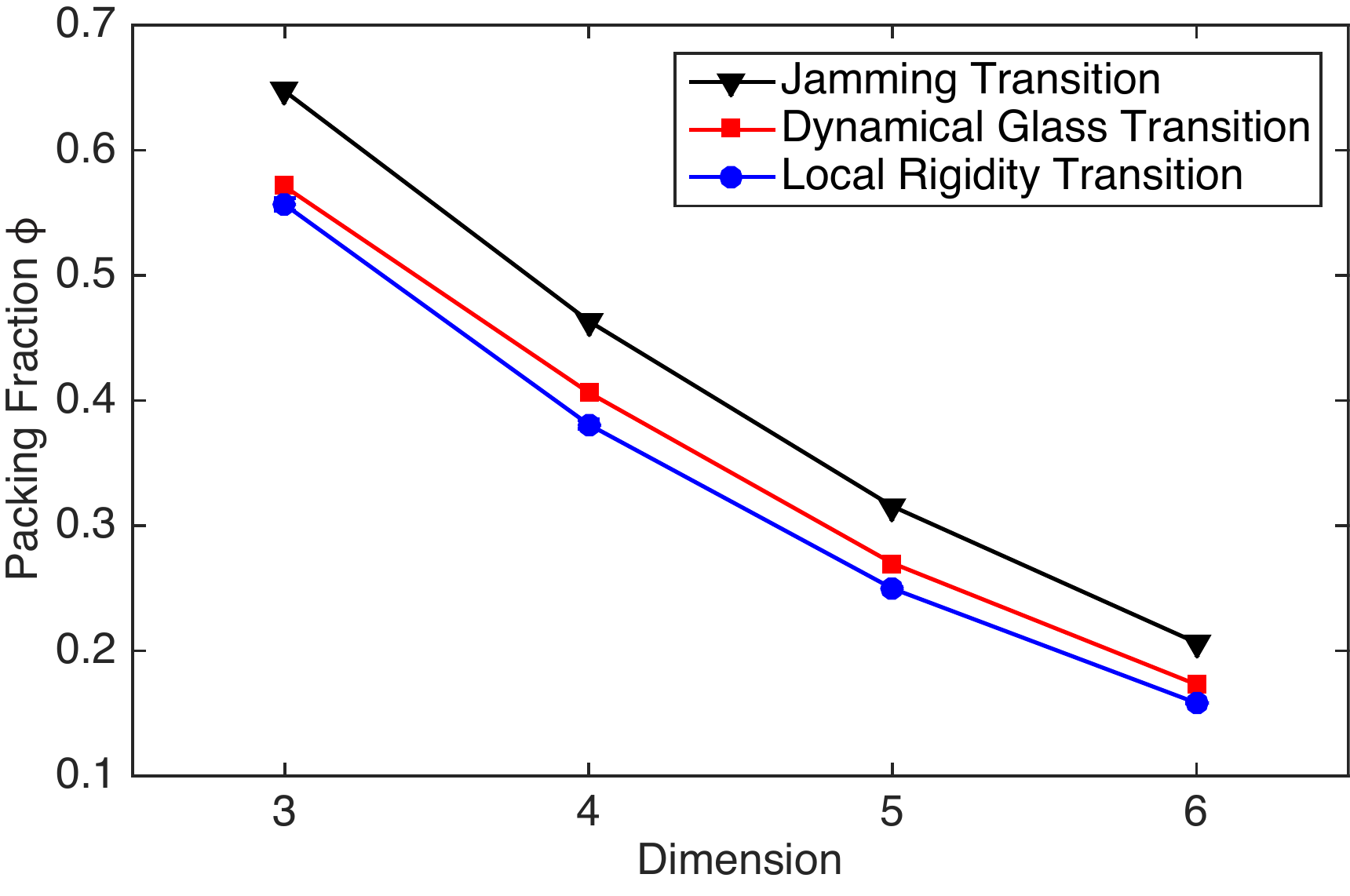}
\caption{Values of jamming point $\phi_\textrm{J}$  \cite{morse_geometric_2014}, the dynamical glass transition point $\phi_\textrm{d}$ in $d=3$ \cite{charbonneau_glass_2011} and $d=4-6$ \cite{charbonneau_dimensional_2012}, and local rigidity onset point $\phi_\textrm{LR}$ in systems with 16384 particles. $\phi_\textrm{LR}$ values have their error in the least significant digit, which is quoted in parentheses.}
\label{fig:phiGphiLR}
\end{figure}

Figure \ref{fig:phiLRAll}a shows the average fraction of locally rigid particles as a function of scaled density in dimensions 3 through 6. In all dimensions, this is nearly zero for low density systems with an increase well below the jamming transition. While in $d$=3, this increase is quite smooth, the transition becomes sharper in higher dimensions. We define $\phi_\textrm{LR}$ as the density at which a locally rigid cluster percolates the system, shown in the inset of Figure \ref{fig:phiLRAll}a. To define percolation, we first identify the largest connected group of locally rigid particles and then apply a burning algorithm to determine whether it percolates through our periodic boundary conditions in all directions. To get the precise value, we average over 8 systems and fit to an error function, which gives both the average value and the standard deviation of the percolation value. Plots of the error function fit are shown in the supplement. We have scaled $\phi$ by $(\phi - \phi_\textrm{LR}) / (\phi_\textrm{J} - \phi_\textrm{LR})$ such that the local rigidity transition is at a scaled density of 0 and jamming is at a scaled density of 1. We find these locally rigid clusters to be predominantly composed of a single giant connected component which is not compact, but rather filamentary in nature. Thus a close inspection of Figure \ref{fig:phiLRAll}a shows that percolation happens almost immediately after the first locally rigid clusters form in $d \ge 4$. Numerical values for $\phi_\textrm{LR}$ for systems with $N=16384$ are given in figure \ref{fig:phiGphiLR}.

Figure \ref{fig:phiLRAll}b shows the effect of varied system size in $d$ = 4, which is used because it is the lowest dimension to show a sharp transition. We find that smaller systems have a gradual transition, while large systems have a very sharp transition. The definition of $\phi_\textrm{LR}$ is robust to different initial conditions, and so we take averages over multiple data sets as described in the caption. The inset shows the evolution of $\phi_\textrm{LR}$ taken from the error function fit as a function of system size. Unlike jamming \cite{goodrich_finite-size_2012}, this measure of the transition point doesn't appear to have any finite size scaling within the error bars reported. Meanwhile, in $d=3$ (which we show in the supplement) there is a significant rise in the value of $\phi_\textrm{LR}$ for large system sizes, suggesting that we are still in a regime dominated by finite size effects and would require truly enormous systems to pin down an asymptotic value with any degree of precision. It is for this reason that we focus on $d = 4$.

The existence of locally rigid clusters depends sensitively on the definition of contacts. Because we are below jamming, all particle overlaps are zero to within quadruple precision and the distinction is drawn between kissing and non-kissing particles.  However, if we artificially degrade our precision by changing the halting condition in our simulations, we can see the effects of finite precision. We do this by varying the energy per contact below which the system is considered minimized as shown in Figure \ref{fig:phiLRAll}c. We find that as the precision is degraded, the transition becomes more gradual. The inset shows that $\phi_\textrm{LR}$ increases to an asymptotic value as we increase precision, with $-\log_{10}\textrm{E}_\textrm{cut} = 60$ corresponding to quadruple point precision. At high energy cutoffs, the system is less able to find fine features in the energy landscape, corresponding to lower energy states. Thus a system with a lower cutoff is able to more finely search for a minimum, allowing for a higher local rigidity point. This is exactly analogous to varying the quench rate in a thermal system. With a slower quench, the system is able to explore a larger portion of phase space and potentially find rearrangements, allowing for a higher glass transition \cite{cavagna_supercooled_2009}. Again, $d=3$ (shown in the supplement) differs from higher dimensions, in that at any reasonable system size $\phi_\textrm{LR}$ does not have energy cut dependence.

% \section{Discussion}

\begin{figure}
\includegraphics[width=1\linewidth]{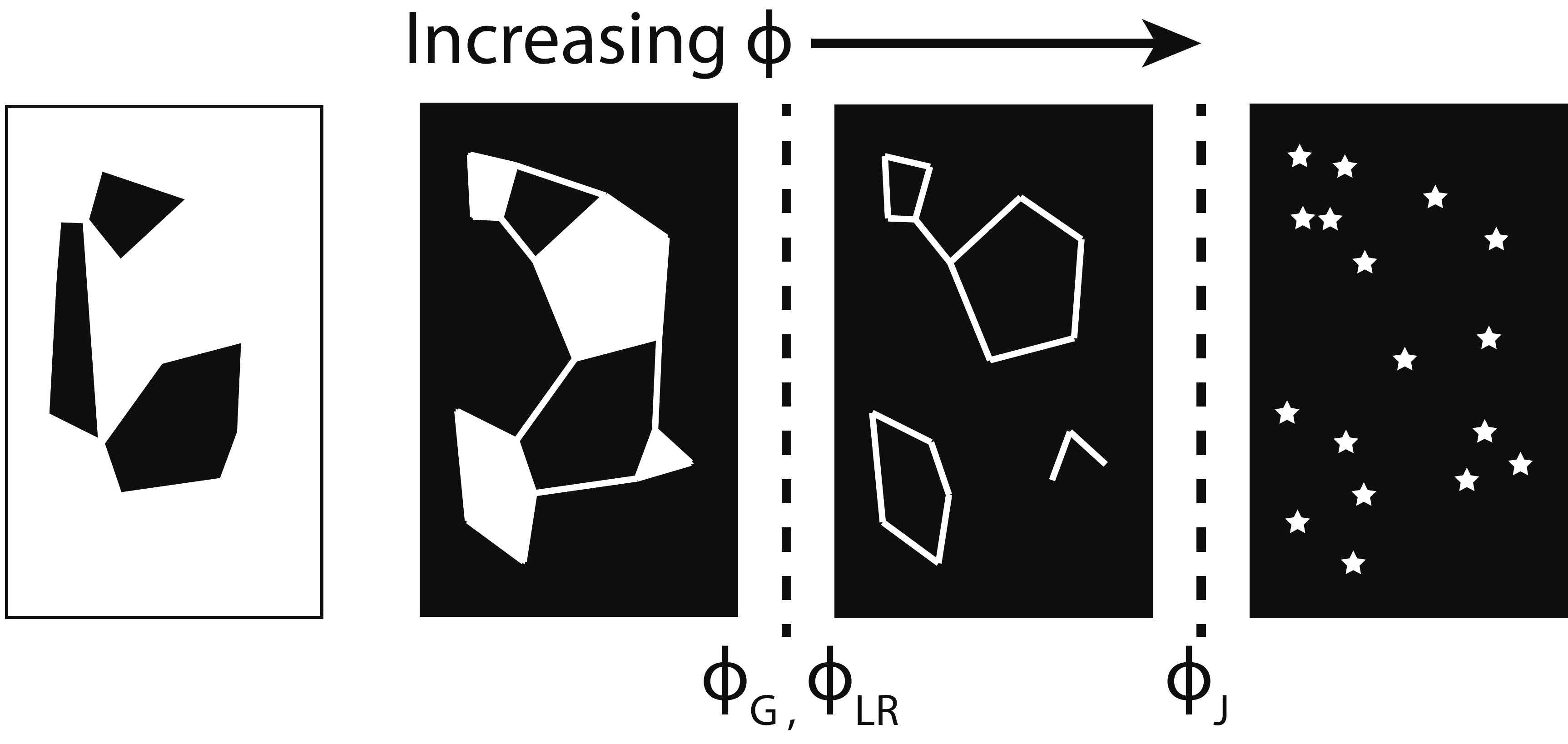}
\caption{A simplified illustration of the allowed configuration space shared by thermal hard spheres and energy minimized athermal soft spheres. White areas represent allowed configurations and black areas represent forbidden configurations. Energy minimized soft sphere athermal systems will generically sit on the boundaries between allowed and forbidden states, while thermal hard sphere systems will explore allowed states. As packing fraction is increased, the volume and dimensionality of allowed configurations decrease, undergoing two distinct transitions at $\phi_\textrm{d} \simeq \phi_\textrm{LR}$ and $\phi_\textrm{J}$.}
\label{fig:phaseSpace}
\end{figure}

This observed local rigidity transition naturally relates to the underlying landscape of allowed configurations. As illustrated in figure \ref{fig:phaseSpace}, thermal systems can be thought of as uniformly spread throughout the volume of the allowed configuration space while our athermal systems generically sit at the interface between allowed and disallowed states, which can be thought of as the shore of the lake of allowed configurations.

For athermal systems this shore of the lake is the only meaningful place at which local rigidity can be measured.  If we attempt to measure $\phi_\textrm{LR}$ on the land (i.e. for an incompletely minimized state), then the results are meaningless as they don't represent a state that is accessible to a hard sphere glass.  If instead we measure $\phi_\textrm{LR}$ in the lake, then the question is ill posed as there is always the freedom to move within this zero energy region to a state in which there are precisely zero contacts. The ``shore of the lake'' is meaningful because it represents the interface between the soft sphere zero temperature states and the hard sphere thermal states.

At low densities most motions are possible; the allowed phase space is dominated by lakes with islands of forbidden configurations. It is easy to find a path from any one configuration to any other and both individual and collective motions are unconstrained and the system is ergodic. The volume and dimensionality of allowed configurations decreases with increasing packing fraction as kissing contacts are formed, undergoing two distinct transitions. The first transition corresponds with these lakes closing into rivers, allowing only collective motions in a few directions (ignoring rattlers), corresponding with soft modes in the thermal glass \cite{lubchenko_origin_2003}. In a thermal system this is the Mode Coupling or Dynamic Glass transition seen at a density $\phi_\textrm{d}$.  In athermal systems this describes our newly introduced local rigidity transition at $\phi_\textrm{LR}$. The second transition occurs when all motion (save that of rattlers) becomes constrained and the allowed space of configurations shrinks to disconnected points.  This is the jamming transition at $\phi_\textrm{J}$.

Figure \ref{fig:phiGphiLR} shows $\phi_\textrm{LR}$, $\phi_\textrm{d}$, and $\phi_\textrm{d}$ as a function of dimension. We observe that $\phi_\textrm{LR}$ precedes the values of $\phi_\textrm{d}$ measured by Charbonneau \textit{et al.} \cite{charbonneau_glass_2011, charbonneau_dimensional_2012}). The mode coupling transition in thermal glasses is the result of the nonlinear feedback mechanisms in the microscopic dynamics of particles becoming so strong that they lead to the structural arrest of the system \cite{kob_mode-coupling_1997, charbonneau_glass_2017}.  Such a situation will only be obtained when the individual and independent motions of particles are tightly constrained and the only free directions remaining in configuration space are collective; precisely the condition needed for local rigidity. Based on the above argument, in the thermodynamic limit and the limit of infinite precision we would expect these two transitions will coincide. To determine whether this measured discrepancy is meaningful will require more intensive numerical study. 

The dynamical transition is marked by a plateau at long times in the mean squared displacement. This plateau is attributed to the caging of individual particles and can also be understood in terms of the underlying landscape revealed by local rigidity in a similar fashion to work done by Wang, Ma, and Stratt \cite{wang_global_2007, ma_potential_2014}. A system of $N$ particles in dimension $d$ will have an $Nd$ diminsional configuration space and thus an expected diffusion constant in that configuration space set by the scale $Nd$. At low packing fractions when motion is unconstrained, the system is allowed to move freely. However, if a fraction $\alpha$ of the available directions in phase space are restricted to a microscopic length scale, as observed by the appearance of local rigidity in the athermal system, then the diffusion constant will drop to $(1-\alpha)Nd$, which restricts motion along smaller channels. Thus, when these channels become small they create the long plateau in the mean-squared displacement and give rise to the onset of rigidity in the dynamical transition.

% \section{Conclusion}

The origin of rigidity in both thermal and athermal systems is controlled by a topological change in the shared energy landscape which prevents individual motions and only allows highly collective motion. Thus the dynamical glass transition is inextricably linked to an athermal local rigidity transition. In this way, the energy landscape becomes a concrete tool instead of merely a suggestive concept. By removing the kinetics of thermal systems we can more plainly see that underlying structural changes in phase space the behavior of both systems.

\begin{acknowledgments}
We thank Paddy Royall for helpful discussion. This work was supported by the NSF under Career Award DMR-1255370 and a Grant from the Simons Foundation No. 454939. The ACISS supercomputer is supported under a Major Research Instrumentation grant, Office of Cyber Infrastructure, OCI-0960354.

\end{acknowledgments}

\bibliography{LocalRigidity}

%\pagebreak
%\vspace{1cm}

\end{document}